\documentstyle[prl,aps,epsfig]{revtex}
\begin{document}

\def\gsim{\;\rlap{\lower 2.5pt
 \hbox{$\sim$}}\raise 1.5pt\hbox{$>$}\;}
\def\lsim{\;\rlap{\lower 2.5pt
   \hbox{$\sim$}}\raise 1.5pt\hbox{$<$}\;}
\def\msol{{\rm\,M_\odot}}
\def\yr{{\rm\,yr}}
\def\kpc{{\rm\,kpc}}
\def\odb{{\eta_{\rm dm}}}
\def\kms{\rm\,km\,s^{-1}}
\def\mpc{{\rm\,mpc}}
\def\au{{\rm\,AU}}
\def\del{{\partial}}
\def\gm{{\rm\,g}}
\def\cm{{\rm\,cm}}
\def\sec{{\rm\,s}}
\def\erg{{\rm\,erg}}
\def\kev{{\rm\,keV}}
\def\mic{{\,\mu{\rm m}}}
\def\ev{{\rm\,eV}}
\def\kms{{\rm\,km\,s^{-1}}}
\def\K{{\rm\,K}}
\def\zsol{{\,Z_\odot}}

\twocolumn[\hsize\textwidth\columnwidth\hsize\csname@twocolumnfalse%
\endcsname
\vspace{3mm}

\draft
\title{The Cold Big-Bang Cosmology as a 
Counter-example to Several Anthropic Arguments}

\author{Anthony Aguirre}

\address{School of Natural Sciences, Institute for Advanced Study, 
Princeton, New Jersey 08540, USA}

\maketitle

\begin{abstract}
A general Friedmann big-bang cosmology can be specified by fixing a
half-dozen cosmological parameters such as the photon-to-baryon ratio
$\eta_\gamma$, the cosmological constant $\Lambda$, the curvature
scale $R$, and the amplitude $Q$ of (assumed scale-invariant)
primordial density fluctuations.  There is currently no established
theory as to why these parameters take the particular values we deduce
from observations.  This has led to proposed `anthropic' explanations
for the observed value of each parameter, as the {\em only} value
capable of generating a universe that can host intelligent life.  In
this paper, I explicitly show that the requirement that the universe
generates sun-like stars with planets does {\em not} fix these
parameters, by developing a class of cosmologies (based on the
classical `cold big-bang' model) in which some or all of the
cosmological parameters differ by orders of magnitude from the values
they assume in the standard hot big-bang cosmology, without precluding
in any obvious way the existence of intelligent life.  I also give a
careful discussion of the structure and context of anthropic arguments
in cosmology, and point out some implications of the cold big-bang
model's existence for anthropic arguments concerning specific
parameters.

\end{abstract}
]

\vskip0.3in
\section{Introduction}

	Current fundamental physical theories and cosmological models
incorporate a number of `parameters' or `constants' which could
theoretically assume different values while leaving the mathematical
structure of those theories unchanged.  Responses to the question of
why any given parameter/constant assumes the particular value it does
fall into three rough categories:
\begin{enumerate}
\item{Like the laws of physics themselves, the value of the
parameter/constant in question is fundamental, and simply a part of
the nature of the universe itself.  Whether or not it `could have'
been different is unclear but irrelevant, since it assumes a unique
value that is fixed in space and time.}
\item{The parameter/constant can ultimately be derived from a fundamental
physical theory with no free parameters.  Thus the observed
`parameter' could not have been different, as it is a purely
mathematical objects.}
\item{The parameter varies between members of a spatial/temporal
ensemble, of which the region of the universe we observe is part of
one member.  The parameter not only could have been different but
{\em is} different in other ensemble members, and the value we
observe depends upon which ensemble member we happen to
inhabit.}
\end{enumerate}

Explanations of the third type have raised considerable interest in
cosmology for two related reasons.  First, a number of inflationary
cosmology and quantum cosmology theories explicitly fail to predict
unique values for cosmological parameters such as the photon-to-baryon
ratio $\eta_\gamma$, the cosmological constant $\Lambda$ or the
amplitude of density perturbations $Q$; rather, these theories yield
only a probability distribution for the parameters, which take
different values in causally disconnected
`sub-universes'~\cite{lindebook,reesbook}.  Second, the observed
values of some parameters/constants (such as the cosmological
constant~\cite{weinberg}) seem to require an incredible degree of fine
tuning if they are to admit explanations of the first or second type.

This fine tuning might be avoided by explanations of the third type,
which allow for a decoupling between the `expected' or `natural' value
of a parameter and the value we observe, by incorporating the fact
that some parameter values may preclude the existence of observers
that could measure those values.  These `anthropic' considerations
imply that the probability of measuring a given value for some
parameter is not simply given by the probability distribution of
values assumed by that parameter among members of the ensemble, but is
modified by the probability of observers arising in each member.
Values of parameters that are very different from the `expected'
values might then be explained as being typical values among the set
of values that allow the formation of observers.

Arguments of this sort have been used to explain the observed value of
$\Lambda$~\cite{weinberg,weinberg1,1995MNRAS.274L..73E,1998ApJ...492...29M,th,gv2,glv},
$\eta_\gamma$~\cite{linde2}, $Q$~\cite{1998ApJ...499..526T}, the
curvature scale~\cite{btip,ht,glv2}, and the density ratio of
non-baryonic dark matter to baryons $\odb$~\cite{linde}.  Generally,
these arguments consider the viability of life in a universe in which
one of these parameters assumes a value ten times smaller or larger
than the observed values {\em with all other parameters fixed}.  If
the formation of observers is strongly suppressed in each alternative
universe, and if the {\em a priori} distribution of parameter values
is fairly flat, it is claimed that the observed value has been
explained.  For example, Tegmark \& Rees~\cite{1998ApJ...499..526T}
(hereafter TR) explain the observed value of $Q\sim 10^{-5}$ by
showing that galaxies could not cool sufficiently if $ Q\lsim
10^{-6}$, and would be so dense as to disrupt most planetary orbits if
$Q \gsim10^{-4}$.

	Two substantial worries arise with respect to such arguments.
First, a parameter value differing by many orders of magnitude (rather than
just one) may correspond to qualitatively different physical processes 
which allow a rather different universe in which life could still arise.
Second, if more than one parameter is to be
explained anthropically, then several parameters must be varied at
once, and there is a risk that degeneracies will occur in which
changing one constant counteracts the adverse effect of changing
another.  This paper argues that these two worries are substantial and
serious, by developing a specific set of cosmological models in which
one or more of the basic cosmological parameters can be altered by
many orders of magnitude, without preventing the formation of
observers (conservatively assumed to be similar to us) in any obvious
way.

	To develop this argument, I first discuss in some detail the
logical structure of the `weak' anthropic arguments in question.  I
then discuss specific contexts (inflation, quantum cosmology, etc.) in
which they may arise.  Next I develop a general class of cosmologies
based on the classical `cold big-bang' (CBB) cosmology and argue that
observers like us could plausibly arise in these cosmologies.  I then discuss
several specific anthropic arguments in light of the cold big-bang
cosmology.

\section{the structure of anthropic arguments in cosmology}

	Anthropic arguments concerning fundamental parameters in a
theory generally are invoked to explain the particular measured value
of a given parameter, in relation to the large range of other possible
values that it seems reasonable to expect that the parameter {\em
could} have assumed.  The `weak' form of the anthropic principle (the
only form this paper discusses) explains particular values of some set
of parameters using a two-part argument.  First, the argument requires
that the parameters in question not only {\em could have} been
different, but in fact {\em do} assume a range of values in a
physically realized ensemble of systems, some of which contain
observers capable of measuring the parameter, and all of which are
similar except for variations in the values of the parameters in
question. The second part of the argument is the self-evident
statement that only systems capable of producing observers can have
their parameters measured.  Thus observers will never measure a
parameter to have a value which would preclude the existence of
observers.  (For example, assuming that atoms are required for life to
exist, no living observer will ever measure a value of the electron
charge incompatible with the existence of atoms).

	The argument is simple, but immediately raises the issue of to
{\em which} parameters describing {\em which} physical systems the
argument might be applied.  For the purposes of this paper, let us
categorize parameters into three groups.  First are parameters which
are known to vary in space or time and may be derivable from more
fundamental parameters; for example, the `solar constant' or the
`Hubble constant'.  Second, the constants\footnote{Excepting $\Lambda$
(which is assumed to vary), I will hereafter reserve the term
`constant' for parameters with fixed values.}  used in the
currently-accepted fundamental physical theories (the standard model
of particle physics and general relativity)~\cite{hog2}.  Third, the key
parameters describing the current `standard model' of cosmology
(defined presently).  This study concerns parameters in the third
category, and addresses anthropic arguments in cosmology which attempt
to explain their values.

\subsection{The general argument applied to cosmology}
\label{sec-acos}

I take the cosmological `standard model' to be a
Friedmann-Robertson-Walker (FRW) big-bang cosmology (i.e. the
cosmology generated by solving Einstein's equations assuming
large-scale homogeneity and isotropy), characterized by a spatial
curvature scale ${\cal R}$ (evaluated at the Planck time, in units of
the Planck length), a photon-to-baryon ratio $\eta_\gamma$, an
(electronic\footnote{I assume for simplicity that the other species
have lepton numbers small compared to the electronic leptons.})
lepton-to-baryon ratio $\eta_L$, a ratio $\odb$ of non-baryonic dark
matter to baryonic matter (evaluated when both are nonrelativistic), a
cosmological constant $\Lambda$ (in Planck units), and a
scale-invariant power spectrum of Gaussian primordial density
perturbations with an amplitude $Q$ on the horizon scale.  This
definition restricts the types of cosmologies that can be considered,
but encompasses those for which anthropic arguments have been made in
the literature.

	Anthropic arguments in cosmology have raised new interest due
to the possibility that the cosmological parameters are {\em not}
uniquely derivable from more fundamental considerations, nor `fixed'
as part of the initial conditions of the universe, but vary among
`sub-universes' with probability distribution
$P(\bar\alpha_1,. .,\bar\alpha_N)$.  Here, the set $\{\alpha_i\}$
stands for some subset of $\{{\cal
R},\eta_\gamma,\eta_L,\odb,\Lambda,Q\}$, and
$d^NP(\bar\alpha_1,. .,\bar\alpha_N)/d\bar\alpha_1..d\bar\alpha_N$ is
the differential probability that a randomly chosen
baryon\footnote{Any number which is conserved during the cosmological
expansion could be used here.}  resides in a sub-universe in which the
parameters $\alpha_i (i=1..N)$ take a given set of values in the range
$[\bar\alpha_i,\bar\alpha_i+d\alpha_i]$. (The sorts of systems which
constitute sub-universes are discussed below in
\S~\ref{sec-contexts}.) But $P$ is {\em not} the probability
distribution of observed values of $\alpha_i$, which is instead (by
anthropic reasoning) proportional to ${\cal
P}(\bar\alpha_1,. .,\bar\alpha_N)\equiv
P(\bar\alpha_1,. .,\bar\alpha_N)\xi(\bar\alpha_1,. .,\bar\alpha_N)$,
where $\xi(\bar\alpha_1,. .,\bar\alpha_N)$ is the total number,
integrated over time, of observers per baryon that are capable of
making independent measurements of $\alpha_i$ in a region of space in
which the cosmological parameters are given by $\bar\alpha_i$. One can
straightforwardly normalize ${\cal P}$ to yield a true probability
distribution, provided that $P\xi$ is integrable over the space of
values attained by the $\alpha_i$ in the ensemble. (The method of
defining probabilities I have chosen is probably most similar to that
of Vilenkin~\cite{vil95}; other authors have formulated anthropic
arguments in ways that are similar in spirit but different in detail.)

	The function $P$ will presumably follow from the (currently
unknown) fundamental physics describing the universe as a whole.  In
this paper, I concentrate on $\xi$ (though I draw some conclusions
about $P$); determining $\xi$ requires a criterion for the existence
of an observer and a method by which to calculate the density of such
observers for a given set of $\alpha_i$ and their values
$\bar\alpha_i$.  Determining what sort of physical configurations
could give rise to a being capable of measuring cosmological
parameters is a rather difficult task which I will sidestep by
confining my criteria to those which are obviously essential for the
existence of life similar to humans.  This assumption is conservative
in the present context in that it grants the anthropic argument
maximal predictive power, and is tantamount to assuming that the
frequency of independent human-type observers greatly exceeds that of
all other types of observers the universe may produce.  The specific
criterion adopted here is to require the formation of a main-sequence
star with a moderate fraction of heavy elements such as C, N, O,
etc. The star must burn steadily and without significant disturbance
(e.g., which would disrupt planetary orbits) for more than an
`evolutionary timescale' $\tau_{\rm ev}$; I take $\tau_{\rm ev} =
5\,$Gyr (the single available observation for the timescale on which
observers arise after the formation of a star).  Adopting this
criterion, I set
\begin{equation}
\xi(\bar\alpha_1,. .,\bar\alpha_i)=\int_{Z_{\rm min}}^{Z_{\rm
max}}dZ \int_{\tau_{\rm ev}}^\infty d\tau \int_0^{t_{\rm
max}-\tau_{\rm ev}} dt {dn(t,\tau,Z; \bar\alpha_i)\over
dt\,d\tau\,dZ}.
\label{eq-xieq}
\end{equation}
Here, ${dn(t,\tau,z; \bar\alpha_i)/ dt\,d\tau\,dZ}$ is the
differential formation rate (per baryon) at time $t$ of stars with metallicity $Z$
which will live undisturbed for time $\tau$; $t_{\rm max}$ is the
lifetime\footnote{$t_{\rm max}\rightarrow\infty$
is allowed for sub-universes which do not recollapse,
but I assume that each baryon is incorporated into a finite number of stars.}
 of the sub-universe in question, and $Z_{\rm min}$ and $Z_{\rm max}$
define the range of allowed metallicities.  Thus
$\xi(\bar\alpha_1,. .,\bar\alpha_i)$ is the total number 
of stars per baryon
 with $Z_{\rm min} \lsim Z \lsim Z_{\rm max}$ that live $\gsim
5\,$Gyr in relative isolation, in a sub-universe with cosmological
parameters $\bar\alpha_i$~\cite{misceff}. For example,
in the observable universe $\xi \sim 0.01m_p/\msol$ since about $1\%$
of baryons form single stars a solar mass or less.  

	Having defined the ingredients, I now discuss the hope of what
I will term the `anthropic program'.  The hope is that given {\em a
priori} calculations of
$d^NP(\bar\alpha_1,. .,\bar\alpha_N)/d\bar\alpha_1..d\bar\alpha_N$ and
$\xi(\bar\alpha_1,. .,\bar\alpha_N)$, their product $d^N{\cal
P}(\bar\alpha_1,. .,\bar\alpha_N)/d\bar\alpha_1..d\bar\alpha_N$ will
have a well-defined global maximum at some set of parameters
$\bar\alpha_i^{\rm max}$.  For each parameter $\alpha_k$, one might
then integrate ${\cal P}$ over the other $\alpha_i$s ($k\neq i$) to
obtain a 1-dimensional probability $d{\cal
P}(\bar\alpha_k)/d\bar\alpha_k$.  If the peak surrounding the global
maximum is very sharp, ${\cal P}(\bar\alpha_k)$ can be used to define
a range of $\bar\alpha_k$ containing (say) 99\% of the probability.
If (and only if) each {\em observed} value of $\bar\alpha_k^{\rm obs}$
falls inside its `highly probable' region, and if we {\em assume} that
we observe the value that a typical observer does, then anthropic
argument has explained their values. In this case, for example, the
`natural' value of $\Lambda$, would be reconciled with its observed
value (which if nonzero is many orders of magnitudes smaller).  Note
that this procedure is very different from the calculation of the {\em
conditional} probability ${\cal P}(\alpha_k|\bar\alpha_1^{\rm
obs},..,\bar\alpha_{k-1}^{\rm obs},..,\bar\alpha_{k+1}^{\rm
obs},..,\bar\alpha_N^{\rm obs})$ of measuring a single
$\alpha_k=\bar\alpha_k$ with the other parameters fixed (i.e. by their
observed values).  Making an anthropic argument using such a
conditional probability is only logically consistent under the {\em
assumption} that the anthropic program will be successful, and that
one can look at variations in a single parameter while keeping others
fixed at their maximally probable values (presumed to be close to the
observed values).

	The anthropic program, however, can fail in three clear ways.
First, there may {\em not} be any well-defined maximum: there may be
degeneracies among two or more parameters such that there are
multi-dimensional surfaces of constant ${\cal P}$ spanning a
relatively large region of parameter space.  In this case, anthropic
arguments still do not provide a satisfactory explanation of the
observed value of those parameters in light of the large range allowed
by fundamental physics and anthropic constraints.  Second, the
well-defined global maximum of $d^N{\cal
P}/d\bar\alpha_1..d\bar\alpha_N$ may {\em not} lie near the observed
values $\bar\alpha_i^{\rm obs}$.  This would imply either a flaw in
some element of the computation of ${\cal P}$, or that we live in
region of parameter space with low probability, or that there is
something fundamentally wrong with the whole anthropic approach (e.g.,
that there are {\em not} other sub-universes with different values of
the parameters in question).  Third, ${\cal P}$ may have two or more
well-defined local maxima.  This would not be a problem in {\em
principle} as long as one of the peaks was much higher than the rest.
But a similar and important practical difficulty can arise if $\xi$
has multiple local maxima.  This is because anthropic arguments in the
literature typically make simplifying assumptions regarding $P$, on
the grounds that $\xi$ must have a peak with a width which is very
narrow compared to the scale over which $P$ varies
(e.g.,~\cite{1998ApJ...492...29M}).  But if multiple peaks (even
narrow ones) occur in widely separated regions of parameter space, $P$
becomes crucial and the implications of computations based on
anthropic reasoning become more ambiguous.

	This paper argues that multiple regions of large $\xi$ {\em
do}, in fact, exist in the set of parameters
$\alpha_i = \{\eta_\gamma,Q,\odb,\eta_L,{\cal R},\Lambda\}$, by
providing a cosmological model in which many of these parameters can
take quite different values than those we observe, without
preventing (according to the criteria define above) the existence of
observers.  Before developing this cosmology and its implications for
the anthropic arguments, I will first outline the cosmological contexts in which
anthropic arguments are typically made.

\subsection{Contexts for cosmological anthropic arguments}
\label{sec-contexts}

	Any attempt to implement the anthropic program described above
requires that the `universe' (i.e. everything that exists throughout
all time) contains an ensemble\footnote{This is {\em not} an ensemble
of imagined identical systems in the Gibbs-Einstein sense, but rather
a canonical or micro-canonical ensemble of weakly or non-interacting
subsystems of a large or infinite physical system.}  of regions which
may be treated as individual FRW cosmologies.  A number of such
`meta-cosmologies' have been proposed.  For example, the `oscillating
universe' model consists of a series of finite-volume, finite-age
cosmologies (e.g.,~\cite{1995MNRAS.275..850B}).  Here, each `big crunch'
is followed by a new `big bang' in which the parameters $\alpha_i$
might be newly drawn from the probability distribution
$P(\bar\alpha_1,. .,\bar\alpha_N)/N_b(\bar\alpha_1,. .,\bar\alpha_N)$, where
$N_b(\bar\alpha_1,. .,\bar\alpha_N)$ is the total number of baryons in
a cosmology with parameters $\bar\alpha_i$.  Then
$\xi(\bar\alpha_1,. .,\bar\alpha_N)$ can be straightforwardly taken to
be the number of stars which form (with $Z_{\rm min} \lsim Z \lsim Z_{\rm
max}$ and lifetime $>\tau_{\rm ev}$) in the cosmology with parameters
$\bar\alpha_i$, divided by $N_b$.  Current astronomical data weighs
strongly against a closed, recollapsing cosmology, so this context is
of value primarily because it is fairly unambiguous.
	
	The regions constituting members of the ensemble could be
separated in space rather than time.  For example, an infinite (or
extremely large) universe in which the parameters $\alpha_i$ vary
spatially could be partitioned into finite regions\footnote{The volume
of space with a given set of $\alpha_i$ need not be finite, because
the ensemble may contain an infinite number of finite regions with the
same $\alpha_i$. I assume that `edge effects' associated with the
boundaries separating these regions are not important.} of differing
$\bar\alpha_i$ which are uniform enough to be treated as individual
FRW cosmologies with the same initial time, each with a fixed number
of baryons.  The relative numbers of regions described by FRW
cosmologies with parameters $\bar\alpha_i^1$ and $\bar\alpha_i^2$
would then give $P(\bar\alpha_i^1)/P(\bar\alpha_i^2)$, and
probabilities as described in \S~\ref{sec-acos} can be defined
unambiguously as long as the universe can be coordinatized in such a
way that there is a time after which no stars form.  Open or critical
globally FRW universes with small density inhomogeneities would (for
example) satisfy these criteria.

	Anthropic arguments can also be made in quantum cosmology
(e.g.,~\cite{th,gv2}).  Here the universe begins in a superposition of
states which `decoheres' into an ensemble of classical cosmologies
with different properties.  The hope is that for a compelling initial
condition, the wave function can be represented as a superposition of
(or at least dominated by) FRW-type cosmologies with different values
of $\alpha_i$.  If these cosmologies were closed, then the situation
would closely resemble the first example of the `oscillating
universe', and the probability
$P(\bar\alpha_1,. .,\bar\alpha_N)/N_b(\bar\alpha_1,. .,\bar\alpha_N)$
of a component of the ensemble being described by parameters
$\bar\alpha_i$ would be proportional to the square of the amplitude of
the term corresponding to that cosmology in the initial superposition.
Note, however, that it is not entirely clear how to extract
probabilities if the superposition contains both open and closed
cosmologies, nor is it clear that the squared amplitudes can be
straightforwardly interpreted as relative frequencies in an ensemble
of classical cosmologies. But assuming that these problems are not
fatal, quantum cosmology does provide a possible framework for
anthropic arguments.

	Applying anthropic arguments to FRW cosmologies embedded in an
arbitrary global geometry is much more difficult because there may not
be a unique globally defined initial time at which to begin the
integration of Eq.~\ref{eq-xieq}.  This is the case in models of
`eternal inflation' in which inflation does not end globally (which
would provide an initial time for the subsequent FRW cosmology), but
always continues in some regions.  The global structure of the
universe approaches an ensemble of thermalized regions separated by
inflating regions~\cite{guth}, and the values of cosmological
parameters describing each region can vary throughout the ensemble.
This is a natural context for anthropic arguments, but as discussed at
length in Ref.~\cite{vvw} (see
also~\cite{vil_lm}), it is a subtle matter to unambiguously define
probabilities in such cosmologies because the probabilities for many
proposed schemes depend strongly upon the coordinate choice: depending
on this choice, a $t=0$ hypersurface can intersect one, many, or no
thermalized regions having different parameter values.  Vanchurin et
al.~\cite{vvw} propose a scheme that apparently circumvents this
problem and gives unambiguous probabilities by calculating
probabilities within any one thermalized region.  For this to work the
parameters must vary continuously in such a way that there is a finite
range of values over which the probability has nonzero measure, so
that the (arbitrarily) chosen thermalized region will contain many
sub-regions with different parameters spanning that range.  Ref.~\cite{gv}
extend this scheme to compute probabilities for
situations in which parameters take on different discrete values in
different thermalized regions.

\section{The Cold Big-Bang Cosmology}
\label{sec-cbb}

	As described above, modern cosmology provides several
plausible (if speculative) contexts in which the universe could
consist of an ensemble of regions, each describable as an FRW
cosmology with different initial conditions.  The thesis of the
present study is that regions with quite different parameters may
support life, thereby greatly complicating or invalidating several
anthropic arguments. I support this thesis by developing a big-bang
type cosmology in which some or all of the parameters
$\eta_\gamma,Q,\odb,\eta_L,{\cal R}$, and $\Lambda$ may differ by at
least several orders of magnitude from the current `standard model' of
cosmology in which they take approximate values of $|{\cal
R}|\approx2\times10^{29}(1-\Omega-\Omega_\Lambda)^{1/2}$, $\eta_\gamma \approx
2\times 10^{9}$, $\eta_L\sim1$~\cite{bg}, $\odb \approx 5-10$,
$\Lambda \lsim 3\times10^{-122}$ (i.e. $\Omega_\Lambda \lsim 0.7$),
and $Q \sim 10^{-5}$.

\subsection{Initial conditions}

	Consider an FRW cosmology with physical baryon number
density $N(t_I)$ at some initial time $t_I$ after which the comoving
baryon number is conserved, and at which the photon-to-baryon ratio is
small ($\eta_\gamma \lsim 10$) relative to that we currently observe.
Choosing $N(t_I) = 10^{35}{\rm cm^{-3}}$ ensures also that nucleons
are nonrelativistic, that known baryonic species other than nucleons have
decayed, and that for $\eta_\gamma \gsim0.01$ a state of nuclear
statistical equilibrium holds~\cite{kauf,1999ApJ...521...17A}.  The
expansion is dominated by relativistic matter for
\begin{equation}
N_I \gsim
10^{39}(\eta_L^{4/3}+\eta_\gamma^{4/3})^{-3}{\,\rm cm^{-3}}
\end{equation}
 or cosmic time 
\begin{equation}
t \lsim 
10^{-4}(\eta_L^{4/3}+\eta_\gamma^{4/3})^{2}\,s.
\label{eq-teq}
\end{equation}
A cosmology with
$\eta_L\sim\eta_\gamma\sim1$ could result from efficient baryogenesis
after an inflationary epoch, or might simply be `assumed' as the
initial state for an FRW cosmology (see \S~\ref{sec-etagamma}).
Cosmic expansion dominated by nonrelativistic matter will steadily
decrease $\eta_\gamma$ (though it is constant for $\eta_\gamma \gg
1$), while non-equilibrium processes may increase it
slightly~\cite{kauf,1999ApJ...521...17A,1982ApJ...252..418H}.

	I leave the rest of the cosmological parameters relatively
unconstrained, except for generally assuming $\eta_L \sim \eta_\gamma$
and $\odb \lsim 100$ (both for convenience) and $Q \ll 1$ (to avoid
complications involved with significant primordial black hole
formation).  I also assume that ${\cal R}$ and $\Lambda^{-1}$ are
large enough for the expansion to be radiation- or

\begin{figure}
\epsfig{file=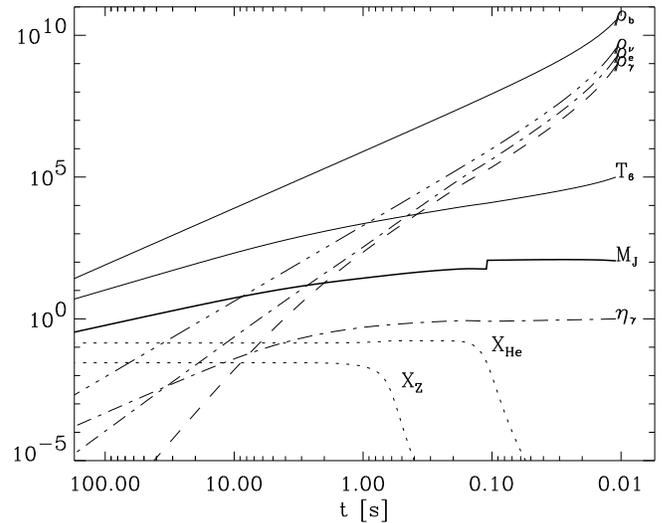}
\caption{
Early evolution of a sample FRW cosmology
with (initial) $\eta_\gamma=1.0$, $\eta_L=2.5$ and $\odb \ll 1$.  
Baryon, photon, electron and neutrino densities
are given in ${\rm g\,cm^{-3}}$.  The Jeans mass is in solar units, the
temperature is in units of $10^6\,$K, and $X_{\rm He}$ and $X_Z$ indicate
mass fractions of baryons bound into Helium and into elements heavier
than Lithium.
\label{fig-thmex}}
\end{figure}

\noindent matter-dominated,
until explicitly stated otherwise (see
\S\S~\ref{sec-lambda},\ref{sec-omega}).

\subsection{Nucleogenesis}

	The evolution of a big-bang cosmology with these initial
conditions through the epoch of nucleogenesis is described in detail
in Ref.~\cite{1999ApJ...521...17A}.  If $\eta_\gamma \gsim0.01$ at
$N\sim10^{35}{\rm cm^{-3}}$, the medium will be hot enough to reach
nuclear statistical equilibrium, and will be dominated by free
neutrons and protons (their ratio depending upon $\eta_L$).  In this
case a standard nucleogenesis calculation, generalized to treat
degenerate leptons, yields the products of primordial nucleogenesis at
late times~\cite{1999ApJ...521...17A}.  For certain combinations of
$\eta_\gamma$ and $\eta_L$, nucleogenesis yields a helium fraction of
25\%, just as in the standard HBB.  However, for $\eta_\gamma \lsim
10$, nucleogenesis also produces heavier elements (metals), yielding
metallicity of $Z \gsim 0.1\zsol$ for a 25\% helium yield.  By varying
$\eta_\gamma$ and $\eta_L$, almost any desired yield of primordial
helium and metals can be obtained.  As a particular example, for
$\eta_\gamma \approx 1$ and $\eta_L \approx 2.5$, the cosmic medium
would emerge with $\sim 15\%$ helium by mass, and $\sim$ solar
metallicity (see Figure~\ref{fig-thmex}). Thus the cosmic medium in a
CBB cosmology can start out with the same level of enrichment as
gas in the HBB which has been processed by stars. (Of
course the ratios of different heavy elements will be different in the
CBB, but C, N and O would be produced in abundance).  Primordial metal
synthesis is suppressed by either high $\eta_L$ or high $\eta_\gamma$.

\subsection{Initial perturbations}

	Whereas in the standard HBB model structure formation cannot
begin until the time of matter-radiation equality ($\sim 10^{12}\,$s),
significantly lower $\eta_\gamma$ allows much earlier and more
efficient structure formation.  In making the present argument I have
assumed that there are scale-invariant primordial density
perturbations of amplitude $Q$ on the horizon scale.  It is
interesting to note that structure may form in CBB models even {\em
without} primordial perturbations (i.e., $Q=0$), because of phase
transitions either in the QCD era~\cite{1982ApJ...252..418H} or (if
$\eta_\gamma \ll 1$) later as the cosmic medium approaches the density
of solid hydrogen~\cite{zel,1973ApJ...179..361L}; in either case the
cosmic medium shatters into `chunks' with random velocities which
induce density perturbations of the form $\delta M/M = (M/m)^{-7/6}$
on a mass scale $M$, where $m$ is the chunk mass~\cite{zel2,peeb}.  If
these chunks survive they can directly coagulate into the first
structures, which heirarchically generate larger
ones~\cite{1973ApJ...179..361L}; if the chunks dissipate they leave
behind the density fluctuations which can see later structure
formation~\cite{1982ApJ...252..418H}.  The former case would lead to
qualitatively different formation of the first structures; in the latter
case the general picture of early structure formation would be affected
quantitatively but not qualitatively.  For simplicity and continuity
with the HBB case, and to investigate general values of $\eta_\gamma$
I will focus on the case of `primordial' scale-invariant perturbations
with $\delta M/M \propto M^{-2/3}$.

\subsection{Structure formation}

	The early history of star formation in a cold FRW cosmology of
the described type has been worked out in some detail by
Carr~\cite{1977A&A....60...13C}.  The key point is that the Jeans mass
when it first falls below the mass enclosed by the horizon, $M_J^i$,
is $\sim (2-10) (\eta_L+\eta_\gamma)^2\msol$ (Hogan 1982) at $t\sim
10^4\,$s, rather than $\sim 10^3\msol$ at $t\sim 10^{13}\,$s in the
HBB.  Stars in a CBB cosmology can therefore begin to form soon after
regions of stellar mass enter the horizon.

Density perturbations on scales smaller than the initial Jeans length
are converted into acoustic waves they enter the horizon and do not
grow subsequently (hence are suppressed relative to larger modes),
therefore the first collapsed regions have mass of order $M_J^i$.
Subsequently, larger and larger regions collapse hierarchically.  Carr
argues that for objects below a critical mass $M_c(Q)$, the cooling
time exceeds the free-fall time, and a single object tends to form;
for $M > M_c$ fragmentation is expected and the collapsing object can
form a cluster.  For $M < M_c$, Carr further argues that the
successive hierarchical collapse of regions will lead to a mass
function of protostellar clumps peaked near $M_c$. (Carr's analysis
gives $dN/dM \propto M^{-1}$ where $N$ is the number density of
collapsing clumps of mass $M$; a Press-Schecter analysis would
give the qualitatively similar $dN/dM \propto M^{-4/3}$.)  For each
clump, $M$ provides an upper limit to the mass of one or more stars
which form from its collapse; for clumps with $M > M_c$, the stellar
mass function depends on the details of the (presumed\footnote{It is not entirely clear
to the author whether or not fragmentation should, in fact, occur in
such systems (see, e.g., the arguments of
Layzer~\cite{1963ApJ...137..351L} and the recent simulations by Abel,
Bryan and Norman~\cite{2000ApJ...540...39A}); but as discussed below
solar mass primordial objects can be obtained without invoking
fragmentation if dark matter of sufficient mass and density exists.})
fragmentation
but objects of $M \ll M_c$ seem likely (since that is what appears to
have occurred in observed globular clusters and galaxies).

	If we accept this basic picture, we may consider in slightly
more detail the early history of the example cosmology plotted in
Figure~\ref{fig-thmex}, with $\eta_\gamma=1.0$, $\eta_L=2.5$ and $\odb
\ll 1$.  At $t < 0.1\,$s, the Jeans mass is constant at
$M_J^i\sim100\msol$ and fluctuations below this mass are strongly suppressed.
  Depending upon $Q_8 \equiv Q/10^{-8}$, two
scenarios then suggest themselves.

\begin{figure}
\epsfig{file=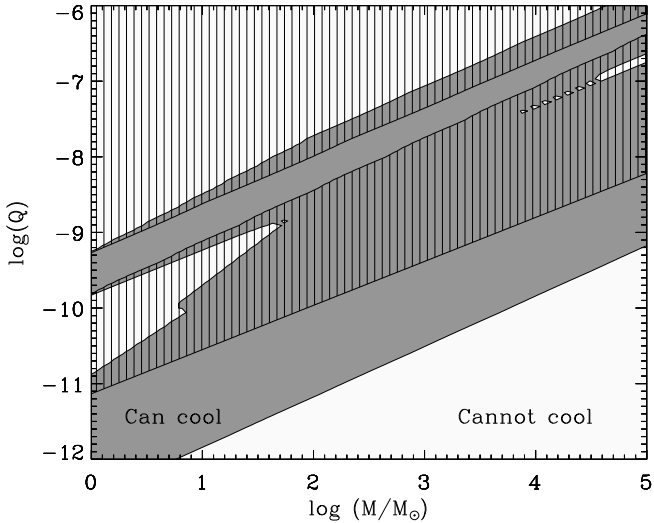}
\caption{
Cooling of primordial clouds in a CBB for
$\eta_\gamma = 1, \odb \ll 1$.  The characteristic temperature $T$ and
density $\rho$ as masses of $100M_{100}\msol$ begin collapse yield an
absorption $\kappa(T,\rho)$ using the fits of 
Bell \& Lin~[31].
Condensations in the region indicated by vertical lines are optically
thick; others are optically thin.  The shaded regions indicate
condensations for which the dynamical time exceeds the cooling time.
}
\label{fig-cool}
\end{figure}

First, if $10^{-3} \le Q_8 \le 1$, the first collapsed regions of
$\sim M_J^i\sim10^2\msol$ cool faster than they collapse (see
Fig.~\ref{fig-cool}), presumably forming stars of much smaller mass.
These primordial groups of stars collapse beginning at time $\sim
10^9Q_8^{-1.5}\,$s, and hierarchical clustering continues, with masses
$\sim 10^{10}Q_8^{1.5}\msol$ collapsing around $t\sim 5$\,Gyr.  In
this scenario, depending on the details of fragmentation, a
substantial fraction of the cosmic baryon mass can form $\sim
1\,\msol$ stars at very early times.

A second, qualitatively different scenario would result from $10
\lsim Q_8 \le 10^4$, in which case masses well above $M_j^i\sim 100\msol$ can
form single systems (see Fig.~\ref{fig-cool}).  The mass function of
the first objects not expected to fragment (by Carr's criterion) will
be dominated by objects of $\approx 2000-5\times 10^6\msol$ which begin
to collapse at cosmic time time $10^9\,$s. In the absence of
significant rotation, those of $\gsim10^5\msol$ should collapse directly
to black holes; the rest should form supermassive stars.  Either type of
object will emit large amounts of radiation at (probably) Eddington
luminosity, converting a fraction $\epsilon$ of its rest mass (or of a
mass of accreted matter similar to its own mass) into energy over a
time $t_e\sim 4\epsilon\times 10^8\,$yr (Rees 1978).  Assuming $t_e$
greatly exceeds the formation time for the objects, this leads to a
photon/baryon ratio of
\begin{equation}
\eta_\gamma \approx 5\epsilon^{5/4}\times 10^{10}.
\end{equation}
Such a large energy release would probably evaporate small structures
and suppress further structure formation until later,
when larger masses go nonlinear and structure can form much as in the
HBB but with the remnants of the initial supermassive objects as dark
matter.  The 

\noindent same basic scenario should occur for any $\eta_\gamma \lsim
50$ and $10 \lsim Q_8 \le 10^4$, though later structure formation
will depend on $Q_8$. 

	The presence of primordial dark matter can modify this picture
substantially, but the effect depends crucially on the mass $m_{\rm
dm}$ of the dark matter particles, and qualitatively on $\odb$.  Dark
matter in a region of a given size can only collapse if the dark
particles free-stream less than that distance during a dynamical time
of the system.  This gives a dark matter `Jeans mass' in terms of the
temperature $T_{\rm dm}$ of the dark matter particles, of $M_J^{\rm
dm}\sim(kT_{\rm dm}/G m_{\rm dm})^{3/2}\rho^{-1/2}$.  Considering only
the dark matter contribution to the density and assuming that the
expansion is roughly adiabatic since the formation of the dark matter,
I find $M_J^{\rm dm} \sim 4(m_{\rm dm}/1\,{\rm GeV})^{-2}\msol$.
Consider the case $\odb \gg 1$. If $M_J^{\rm dm} \ll M_J^b$ (where the
latter is computed neglecting dark matter), then the dark matter will
just decrease the initial (baryonic+dark) Jean mass, $M_J^i$, by a
factor $(1+\odb)^{1/2}$ so that smaller primordial objects can
form. If, however, $M_J^{\rm dm} \gg M_J^b$, then only primordial
objects of mass $M_J^{\rm dm}$ or larger can form, since the Hubble
drag due to the dominant dark matter component prevents the growth of
smaller baryonic perturbations.  Now consider $\odb \ll 1$.  Dark
matter perturbations on a mass scale $M$ with $M_J^{\rm dm} \ll M \ll
M_J^b$ cannot grow because of Hubble drag due to baryons, but will not
free-stream away.  This is important because the baryonic jeans mass
decreases with time, reaching $\sim1\msol$ at $t\sim 30\,$s in our
example cosmology.  Without dark matter fluctuations on this scale
have been suppressed early on; but with dark matter, the small-scale
fluctuations are preserved, and have amplitude $\sim Q\odb/(1+\odb)$.
Thus solar mass objects (or smaller) can still collapse fairly early,
as long as $\odb$ is non-negligible.  Since the collapse time scales
with $M$ but $M_J$ falls slightly more slowly than $1/t$, these
objects will collapse {\em after} the larger objects of $M_J^i$, but
can still survive if they form in regions that are underdense on a
scale $M_J^i$ but overdense on the solar mass scale. 

	Whether the first stars form as $\sim 100\msol$ groups or as
individual Jeans mass objects, they will soon find themselves in a
growing hierarchy of stellar systems, and we must check that any
nascent planetary systems or proto-planetary disks are not disrupted
through stellar encounters.  Using standard linear theory and
spherical collapse, a mass perturbation of mass $100M_{\rm 100}\msol$
turns around at roughly $t\approx 10^9Q_8^{-1.5}M_{100}\,$s, and
virializes at radius $r_V\approx 10^{15}Q_8^{-1}M_{100}\,$cm at
density $\rho_V\approx 5\times 10^{-11}Q_8^{3}M_{100}^{-2}\,{\rm
g\,cm^{-3}}$.  If such a clump fragments into $100M_{\rm
100}/(1+\odb)$ protostars of solar mass, each protostar forms from a volume with
characteristic radius $\sim
2\times10^{14}Q_8^{-1}M_{100}^{2/3}(1+\odb)^{1/3}\,$cm, and can
therefore contract by a factor of ten or more to produce a
protoplanetary disk of $\gsim 1\,$AU for $Q_8 \lsim 1$.  The
protostars will have velocity dispersion $\approx Q^{1/2}c$, and the
system will quickly relax after a time
\begin{equation}
\tau_{\rm rlx}\approx10^9Q_8^{-1.5}M_{100}^2(1+\odb)^{-1}\chi{\rm \,s},
\end{equation}
where $\chi \equiv (\log40)/[\log 40M_{100}/(1+\odb)]$~\cite{bt}.  The
mean time between (proto)stellar encounters with impact parameter $b
\approx 1\,$AU (which would disrupt the formation or orbits of
earth-like planets) is (see~\cite{bt}, p. 541):
\begin{equation}
\tau_{\rm coll} \approx 3.2\log\left({40M_{100}\over 1+\odb}\right)\left({\Theta^2\over1+\Theta}\right)\tau_{\rm rlx},
\label{eq-tdis}
\end{equation}
where 
\begin{equation}
\Theta\equiv{G\msol\over2v^2b} \approx 0.7Q_8^{-1}({\rm AU}/b).
\end{equation}
Thus $\tau_{\rm coll} \ll 5\,$Gyr, which would bode ill for any
forming planets.  However, approximately 1/100th of the cluster's
stars would evaporate each relaxation time~\cite{bt}, so at least some
stars will avoid collisions (by being evaporated) for $Q_8=1$, and the
entire cluster will evaporate before significant collisions occur if
$Q_8 \ll 1$ (or $\odb \gg 1$).  This evaporation occurs as the
relaxation of the cluster moves stars into the high-energy tail of the
Maxwell distribution, and does not require close stellar encounters to
proceed.  After the evaporation (or if formed alone in its halo), a
given (proto)star will likely find itself in a larger mass
condensation; but since $\tau_{\rm coll}/\tau_{\rm evap}$ increases
(logarithmically) with $M$, it cannot experience a planet-disrupting
encounter before this new mass condensation evaporates, and so
on.\footnote{Nearby supernovae might also disrupt protoplanetary
disks.  When the first supernovae explode at $\sim 10^7\,$yr, most
stars would be in clusters of $\sim 10^7Q_8^{3/2}(1+\odb)^{-1}$ stars
with radii $\sim 100Q_8^{1/2}\,$pc, i.e. large globular-cluster type
objects.  It is not presently clear whether or not this might be fatal
to forming planets. Note also that supernovae may also help unbind
clusters by removing gas.}

\subsection{Summary of CBB models}

	In summary, I have argued that for values
of $\eta_L$ and $\eta_\gamma$ of order unity, an FRW cosmology
can begin with solar or greater metallicity and with a Jeans mass $M_J
\lsim 100\msol$ at very early times.  Adding dark matter with mass $\gsim
1$\, GeV yields objects with baryonic mass $M \lsim
100/[\odb(1+\odb)^2]\msol$ and preserves density fluctuations in
$\lsim 1\msol$ regions.  Structure formation depends crucially on the
primordial perturbation amplitude $Q$.  For $10^{-11} \le Q \le
10^{-8}$ and $\odb \ll 1$ the first
collapsed objects are formally unstable to fragmentation.  In this
picture there are three distinct ways solar mass stars can form:
first, by the fragmentation of the first $\sim 100\msol$ objects;
second, as solitary objects in large dark matter halos (if $\odb \gsim
1$); third as solar-mass overdensities embedded in `void' regions,
when the Jeans mass drops to a solar mass.  These stars generally form
beginning at time $\sim 100$\,yr, and should be able to survive
without experiencing encounters which disrupt their planetary systems,
as long as $Q \lsim 10^{-8}$ and/or $\odb \gsim 1$.  For $Q \gsim 10^{-7}$
very massive primordial stars and quasars could form, inhibiting
structure formation until much later, when it would form as in the HBB
(but with Population III remnants as dark matter and with arbitrary
primordial metallicity).

	In the notation introduced in \S~\ref{sec-acos}, these arguments
suggest 
\begin{equation}
\xi(\eta_\gamma\sim 1,10^{-11} \le Q \le 10^{-5})
\sim \xi(\eta_\gamma\sim 10^{9},Q\sim10^{-5}),
\end{equation}
where $\xi$ is the number of solar mass stars per baryon, and where the
parameters not listed can be (but are not necessarily) the same in both
cases.  While the argument for $\msol$ stars in a CBB is not
incontrovertible, it seems doubtful that a much stronger argument for
$\msol$ star formation could be made in the HBB model without the
benefit of {\em observations} of solar mass stars and the assumption
that the HBB model describes the observable universe.

\section{Arguments concerning the entropy per baryon}
\label{sec-etagamma}

	I have outlined a cosmological model with $\eta_\gamma
\sim 1 \ll \eta_\gamma^{\rm obs} \sim 10^9$, which appears to allow
the formation of life-supporting stars.  This serves as a
counterexample to any argument which attempts to rule out cosmologies
with $\eta_\gamma \ll \eta_\gamma^{\rm obs}$ using anthropic
arguments.  It is therefore interesting to discuss how cosmologies
with $\eta_\gamma \sim 1$ might come about, and what arguments have
been forwarded against them.

	A major challenge in cosmology is to understand the origin of
the observed nonzero baryon number, given that almost all models
incorporate an early baryon-nonconserving GUT phase and/or an
inflationary phase, both of which erase baryon number.  One of the
more attractive scenarios for generating baryon number is the
Affleck-Dine (A-D) mechanism~\cite{ad,drt}, which emerges
somewhat naturally from supersymmetric models, and is compatible with
the rather low reheating temperatures that may be required by some
inflationary models~\cite{linde2}.  The simplest versions of this
mechanism, however, tend to generate $\eta_\gamma \sim 1$ rather than
the much greater observed value.  This finding led to a number of
explanations involving either a modification of the theory which
suppresses its efficiency~\cite{drt}, or entropy generation after
baryogenesis~\cite{camp}, or an anthropic argument such as that by
Linde~\cite{linde2}.  I will discuss Linde's argument first, then make
a few comments on the general possibility of low-$\eta_\gamma$
cosmologies since they are a crucial ingredient of the remainder of
the paper.

	In Linde's scenario, the universe is comprised
of a vast number of exponentially large and causally disconnected
regions carrying random values of the field $\tilde\phi$, which
determines the photon-to-baryon ratio after A-D baryogensis.
`Typical' domains with $\tilde\phi \sim \pm m_{\rm plank}$ generate $\eta_\gamma
\sim \pm 1$ but much more rare sub-universes could carry $\eta_\gamma
\sim \pm 10^9$.  Fixing $Q$, Linde argues that $\eta_\gamma \ll
10^9$ would lead to extremely dense galaxies, with (density)$\propto
\eta_\gamma^{-3}$.  This could prevent the survival of planetary
systems (see also TR and ~\S~\ref{sec-q} below) and thus
anthropically limit $\eta_\gamma$ to large values.

	This argument is subverted in two ways by the (theoretical)
existence of the CBB cosmologies outlined in \S~\ref{sec-cbb}.  First,
even if $Q \sim 10^{-5}$, low $\eta_\gamma$ can lead to most of the
early cosmic medium collapsing into Population III objects which
generate entropy and can increase $\eta_\gamma$ to $\sim 10^9$, thus
restoring a semblance of the `standard' picture of galaxy formation at
much later times (see \S~\ref{sec-cbb}
and~\cite{1978Natur.275...35R}. For a discussion of the possibility
that {\em our} universe is of this type,
see~\cite{1999ApJ...521...17A,1982ApJ...252..418H,zel,1973ApJ...179..361L,1977A&A....60...13C,2000ApJ...533....1A,1982ApJ...255..401W}).
Second, if $Q$ varies in addition to $\eta_\gamma$ and $Q \ll 10^{-5}$
is allowed, solar mass stars can (plausibly) form primordially in the
CBB.  While the stars may exist in clusters which are extremely dense,
I have argued that these clusters would evaporate into larger, less
dense structures before stellar collisions destroy planetary systems
around their component stars.  Thus there is no clear obstacle to the
generation of life-supporting stars in cosmologies with
$\eta_\gamma\sim1$, and the anthropic argument cannot explain the
observation of $\eta_\gamma\sim 10^9$.

	A possible objection to the assumptions of this paper is that
it might be difficult to produce cosmologies with $\eta_\gamma\sim1$
and $\eta_L \sim 1$ in the ensemble.  This is because electroweak
`sphaeleron' interactions which violate $B+L$ but preserve $B-L$
(where $B\equiv\eta_\gamma^{-1}$ and $L\equiv\eta_L B$) are in equilibrium at
temperatures $\gsim 100\,$GeV and tend to wash out $B+L$, giving $B/L
< 0$ and requiring $B-L \neq 0$ for there to be any baryons
left~\cite{kt,dr}.  Generation of large $B-L$ is quite possible in
Affleck-Dine baryogenesis, but negative lepton number can cause
problems in the CBB model because abundant antineutrinos lead to
neutron domination during nucleogenesis and hence to a metal-dominated
medium~\cite{1982ApJ...252..418H,1977A&A....60...13C}.  It is unclear
whether or not such a cosmology could support life like our own.
There are, however, at least three possible ways to avoid this
difficulty.  First, Davidson et al.~\cite{dmo} have argued that the
A-D condensate can survive long enough (before decaying into baryons)
to suppress sphaeleron interactions below their critical temperature.
Second, so-called `B-balls' can form from the condensate and protect
the baryon number from erasure~\cite{em}, then later decay into
baryons.  Third, since $B-L$ is preserved family-by-family, it is
possible to obtain $B/L_e > 0$ while $B/L < 0$, by compensating for
the positive electron lepton number (desired for neutron suppression)
using large negative $\tau$ and/or $\mu$ lepton numbers. For example,
setting $L_\mu = B/3$, $L_e=B/3+1$ and $L_\tau=B/3-25/13$ yields,
after $B+L$ erasure, $B=6/13, L=15/13$ and hence $L/B=2.5$ (see
Ref.~\cite{kt}, p. 189 for details).  Thus the desired $B/L$ can be
obtained by tuning $B/3-L_i$ for each family $i$.  This is possible as
long as weak interactions are not fast enough to equalize $L_e,
L_\tau$ and $L_\mu$.

Thus we see that given an ensemble of sub-universes with different
 cosmological parameters, members with $\eta_\gamma \sim 1$ are quite
 possible, and anthropic arguments do not rule out their observation
 even if all other paramters are fixed.  And as discussed below, if
 other parameters vary, values of $\eta_\gamma$ a few orders of
 magnitude different are also allowed.

\section{Arguments concerning the amplitude of primordial fluctuations}
\label{sec-q}

	A second anthropic argument which is directly contradicted by
the claim that a cosmology with $\eta_\gamma\sim1$ and $Q\lsim10^{-8}$
can support life is the argument that only $Q\sim10^{-5}$ is
anthropically allowed (TR).  In essence, TR argue that for $Q \ll
10^{-5}$ structures are too cold and diffuse to cool efficiently into
galaxies, whereas for $Q \gg 10^{-5}$ galaxies are too dense for
planetary systems to survive for 5\,Gyr.  However (and as noted by
TR), limits on $Q$ depend on the other cosmological parameters; for
example, while lower $Q$ impedes the formation of structures that can
cool, lower $\eta_\gamma$ enhances it.  Indeed, the lower limit on $Q$
of TR depends on $\eta_\gamma^{4/3}$ if all other parameters are fixed
(their Eq. 11), so a very low value of $Q$ can be accommodated if
$\eta_\gamma$ is lowered in tandem.

	Against this possibility, TR offer two comments.  First, that
$Q$ must be large enough that the characteristic energy of virialized
structures exceeds the atomic energy scale of $\sim1$\,ryd, lest
cooling be very inefficient.  This gives $Q \gsim 10^{-8}$ (their
Eq. 6).  But this assumes that only atomic transitions can cool the
gas; for the high densities and low temperatures of the first objects
in the example cosmology of \S~\ref{sec-cbb}, molecular vibrational
and rotational cooling, and dust cooling, all with characteristic
energy scales orders of magnitude below atomic energy scales, would be
very efficient, particularly since the objects can have arbitrarily
high metallicity.  TR's second objection is that lowering
$\eta_\gamma$ increases the likelihood of planetary disruptions, the
frequency of which increases roughly as $\eta_\gamma^{-4} Q^{7/2}$
(see TR's Eq. 18 or Eq.~\ref{eq-tdis} of \S~\ref{sec-cbb} given
$M_{100}\propto \eta_\gamma^2$).  Like the cooling constraint, this
allows lower $\eta_\gamma$ in combination with lower $Q$, but with a
different scaling, and the region of $\eta_\gamma$ satisfying both
cooling and disruption constraints vanishes if $Q$ is too low.  But as
argued in \S~\ref{sec-cbb}, for sufficiently low $Q$ or high $\odb$
stellar clusters should always evaporate before planet-disrupting
collisions can occur, removing the disruption constraint and allowing
very low $\eta_\gamma$ and $Q$.

	Thus it seems that while TR have provided plausible arguments
why $Q$ could not vary by one or two orders of magnitude without
suppressing the formation of sun-like stars {\em if} the other
cosmological parameters are fixed at their observed values, their
calculations do allow somewhat different (by $1-2$ orders of
magnitude) values of $Q$ if $\eta_\gamma$ is also somewhat different
(this is another argument against a strict anthropic constraint on
$\eta_\gamma$).  Furthermore, the arguments of this paper indicate
that variations in $Q$ (and $\eta_\gamma$) of {\em many} orders of
magnitude are allowed because qualitatively new physics becomes
important.

\section{Arguments concerning the ratio of baryonic to dark matter}
\label{sec-dmrat}

	While little is known about the nature of the (probably) cold,
dark, (probably) noninteracting dark matter that is postulated as part
of the standard cosmological model, it {\em is} known that it seems to
have $\sim 4-10$ times the mass density of baryons inferred from
primordial nucleogenesis constraints.  For most dark matter
candidates there is no obvious reason why the dark and baryonic matter
densities should be similar, so one might appeal to anthropic
arguments for an explanation.

	Linde~\cite{linde} constructs such an argument to explain why the
axion-to-baryon density ratio could take a value of $\odb \sim 10$
rather than the $\odb \sim 10^8$ expected if the axion field $\phi$ is in
the natural range of $\sim 10^{16}-10^{17}$\,GeV.
Assuming the other cosmological parameters -- and in particular
$\eta_\gamma$ and $Q$ -- to be fixed, Linde considers a hypothetical
sub-universe with $\phi$ ten times its `observed' value, which leads to
$\odb \sim 400-1000$.  By his reasoning, structures
should then be $\sim 10^8$ times more dense than observed galaxies and
thus presumably incapable of supporting intelligent life.

	As discussed in \S~\ref{sec-q}, however, changes in the ratio
$(1+\odb)/\eta_\gamma$ of matter to radiation can be largely
compensated for by changes in $Q$, since the virial density is
$\propto Q^3 (1+\odb)^4/\eta_\gamma^4$ (TR, Eq. 5) and hence the
number density of stars is $\propto Q^3\odb^{-3}\eta_\gamma^{-4}$ (for $\odb
\gg 1$).  So in Linde's example, the increase in density due to the
increase in $\odb$ could be offset entirely by a decrease in $Q$ of a
factor $10^{-2}$. According to TR's analysis the structures would
still be able to cool for this combination of $Q$ and $\odb$ (though
for much higher $\odb$ -- and hence much lower $Q$ -- atomic cooling
would fail).  This weakens Linde's anthropic argument for low $\odb$.
The degeneracy seems also to allow for values of $\odb \lsim 1$, since
the matter-to-photon ratio would only change by a factor of $\lsim 10$.
Note, however, that $\odb \ll 1$ would qualitatively change the HBB
because structure formation may begin to become `top down' since
fluctuations in the baryons below the Jeans mass at matter domination
($\sim 10^{15}\msol$) would be suppressed.

	The CBB model can also form stars when $\odb \ll 1$ if
fragmentation of primordial objects is effective, and it is possible
to construct a CBB model with $\odb \gg 100$ by properly tuning other
parameters.  But because variations about the standard HBB model can
already give cosmologies with $\odb \sim 1000$ or $\odb \ll 1$ that
can form stars efficiently (as long as $Q$ can also vary), the CBB
scenario does not add anything particularly useful.

	Because of the degeneracy between $\odb$ and $Q$, it seems
that the observed value of $\odb$ cannot be explained by anthropic
means unless the probability distribution $P$ is peaked more strongly
toward higher values of $Q$ than toward higher values of $\odb$,
i.e. unless $P(\odb\sim1000,Q\sim10^{-7}) \ll
P(\odb\sim10,Q\sim10^{-5})$ (where here and henceforth
$P(\alpha_k\sim\bar\alpha_k)$ should be interpreted as
$dP(\bar\alpha_k)/d\bar\alpha_k$, integrated over $\alpha_k$ within an order
of magnitude of $\bar\alpha_k$.)  This can be seen either as evidence
against an anthropic argument for $\odb$ or, if preferred, as a
constraint on $P(\odb,Q)$.

\section{Arguments concerning the cosmological constant}
\label{sec-lambda}

	Anthropic arguments have been forwarded a number of times~\cite{weinberg,weinberg1,1995MNRAS.274L..73E,1998ApJ...492...29M,th,gv2,glv} to explain the vast difference between the `natural' value of the
cosmological constant (roughly the inverse Planck length squared,
$l_{\rm pl}^{-2}$), and its small or vanishing observed value of
$\lsim 3\times 10^{-122}l_{\rm pl}^{-2}$.
Because the energy density of clustering matter decays more quickly
than vacuum energy as the medium expands, in any FRW cosmology with
$\Lambda > 0$ structure formation ceases after some time $t(\Lambda;
\bar\alpha_i)$ at which vacuum energy dominates the cosmic energy density
($\alpha_i$ are the cosmological parameters other than $\Lambda$).  It
is argued that this gives an anthropic `upper bound' on $\Lambda$: if
this time occurs before the collapse of structures capable of forming
solar mass stars and planets, no observer like us can measure the
corresponding value of $\Lambda$.  In the notation of this paper, if
the first structures form at cosmic density $\rho_{\rm fs}(\bar\alpha_i)$
in a cosmology with parameters $\bar\alpha_i$, then
$\xi(\Lambda,\bar\alpha_i)=0$ for $\Lambda \gsim 8\pi G\rho_{\rm fs}/c^2$
(see Ref.~\cite{weinberg1} for a more precise criterion).  More
sophisticated versions of this argument attempt to calculate ${\cal
P}(\Lambda;\bar\alpha_i^{\rm obs})$, i.e. the conditional probability of
an observer measuring a value $\Lambda$, with the other $\alpha_i$
fixed at their `observed' values. In these
papers~\cite{1995MNRAS.274L..73E,1998ApJ...492...29M,glv}, ${\rm \cal
P}$ is computed by calculating $\xi(\Lambda; \bar\alpha_i^{\rm obs})$ and
multiplying by an assumed $P(\Lambda)$; they find a probability
function which peaks at $\Lambda$ comparable to -- but somewhat larger
than -- the value indicated by observations.

	It is simple to see how the anthropic upper bound to $\Lambda$
can change if cosmological parameters other than $\Lambda$ vary
between members of the ensemble of sub-universes postulated by the
anthropic program.  As noted, for example, by TR, since $\rho_{\rm
fs}$ varies with the $\alpha_i$ (excluding $\Lambda$), so does the
anthropic upper bound to $\Lambda$.  To make this ambiguity concrete,
consider the hypothetical cosmology with $\eta_\gamma=1.0$,
$\eta_L=2.5$, $Q \lsim 10^{-8}$ and $\odb \lsim 10$ developed in
\S~\ref{sec-cbb}.  In this cosmology, the first (solar mass) stars may
form at time $t_{\rm fs} \sim 10^9Q_8^{-1.5}\,$s when the cosmic
density is $\rho_{\rm fs} \sim 10^{-12}Q_8^{3}\,{\rm g\,cm^{-3}}$.
Once this first generation of stars (or star clusters) forms,
subsequent domination of the cosmic expansion by vacuum energy should
not affect the development of life (as also argued by
Weinberg~\cite{weinberg1}).\footnote{Weinberg also noted the
requirement that the stellar clusters be massive enough to retain
metals generated in supernovae.  This requirement is unnecessary in
the present case because ample metals can be generated primordially.}
Allowing vacuum domination soon after the turnaround of these first
structures yields an anthropic upper bound of $\Lambda \lsim
4Q_8^{3}\times10^{-105}$ (in Planck units), about 17 orders of
magnitude larger than the upper bound on the observed value of
$\Lambda.$ In \S~\ref{sec-cbb} I argued that $ \xi(\eta_\gamma\sim
10^{9},Q \sim 10^{-5})\sim\xi(\eta_\gamma\sim 1,Q \lsim 10^{-9})$,
i.e. similar numbers of life-nurturing stars per baryon might
plausibly arise in the example CBB model and in the HBB.
Weinberg~\cite{weinberg1} has conjectured that $P(\Lambda)=const.$,
therefore {\em if} we assume that independent of $\Lambda$ the two
cosmologies have similar {\em a priori} probability, i.e.,
\begin{equation}
P(\eta_\gamma\sim 10^{9},Q\sim10^{-5}) \sim
P(\eta_\gamma\sim 1,Q \lsim 10^{-9}),
\end{equation}
Then it would be about $10^{17}$ times more likely for an observer to
find themselves in a cold cosmology with an enormous $\Lambda$ than in
a cosmology like the one we observe.  Thus the anthropic explanation
for a small value of $\Lambda$ fails {\em if} the ensemble of
cosmologies comprising the universe contains cosmologies with values
of $Q$ and $\eta_\gamma$ much smaller than those we observe, unless
the $\Lambda-$independent probability of forming those cosmologies is
many orders of magnitude smaller than that of forming standard HBB cosmologies.

\section{Arguments concerning the curvature scale}
\label{sec-omega}

	As for $\Lambda$, anthropic arguments have been invoked to
explain the difference between the `natural' value for the curvature
scale of a Planck length (${\cal R}\sim 1$), and its observed value of
${\cal R} \gsim 10^{29}$. This large difference has been termed the `flatness
problem', and is one of the prime motivations for considering
inflationary models.  Anthropic arguments can, however, still be made
either in the absence of inflation, or within open inflation
models.

	If an arbitrary FRW cosmology is closed, a straightforward anthropic
constraint on ${\cal R}$ arises from the requirement that the
time before the big-crunch must exceed the timescale for the development of
intelligent life $\tau_{\rm ev}\sim5$\,Gyr.  

	For open cosmologies, anthropic arguments quite similar to
those concerning $\Lambda$ have been formulated~\cite{btip,ht,glv2}.  In
simplest form, these arguments require that structures capable of
forming stars and planets form before the cosmic medium becomes
curvature dominated; this gives an anthropic lower limit on ${\cal
R}$.  Curvature domination occurs roughly when
\begin{equation}
{\cal R}[a(t)/a(t_{\rm pl})]l_{\rm pl}\sim\,ct,
\end{equation}
where $t$ is cosmic time, $a(t)$ is the scale factor and $l_{\rm pl}$ and
$t_{\rm pl}$ are the Planck length and time.  This yields
the anthropic constraint
\begin{equation}
{\cal R} \gsim \left({t_{\rm eq} \over t_{\rm fs}}\right)^{2/3}
\left({t_{\rm pl} \over t_{\rm eq}}\right)^{1/2}\left({c\, t_{\rm fs}
\over l_{\rm pl}}\right),
\end{equation}
where $t_{\rm eq}$ is the time when relativistic domination ends (see
Eq.~\ref{eq-teq}), and $t_{\rm fs}$ is the time when the first
star-forming bound structures form.  For the standard cosmological
model with $\eta_\gamma\sim10^9$ and $Q\sim 10^5$, the first
structures form at $t_{\rm fs}\sim10^8-10^9$\,yr and $t_{\rm eq}\sim
10^4\,$yr, giving the constraint ${\cal R} \gsim 10^{29}$.

If we allow very different values of $Q$ and $\eta_\gamma$, then the
constraint weakens considerably.  Considering the example cosmology of
\S~\ref{sec-cbb} in which stars form beginning at $t\sim
10^9Q_8^{-3/2}\,$s and using Eq.~\ref{eq-teq} for $t_{\rm eq}$, the
constraint becomes
\begin{equation}
{\cal R} \gsim 10^{24}Q_8^{-1/2}.
\end{equation}
Thus the curvature scale can be $\sim 10^5$ times lower in a cold
cosmology, and measurement of a small value of ${\cal R}$ and a cold
cosmology will be much more probable than measurement of larger ${\cal
R}$ and a hot cosmology, in any universe in which the {\em a priori}
probability of hot and cold cosmologies is similar and in which the
{\em a priori} probability distribution of ${\cal R}$ is peaked at
small values (the latter being the only situation in which anthropic
arguments about ${\cal R}$ in open cosmologies are useful).

\section{What if $\eta_\gamma$ does not vary?}

	The difficulties for the anthropic program pointed out in this
paper all depend on the assumption that low-$\eta_\gamma$ cosmologies
exist in the ensemble comprising the universe which is posited in any
cosmological anthropic argument.  Can all of the difficulties be
avoided if it is assumed that $\eta_\gamma$ is the same in all
ensemble members?  Perhaps, but this is by no means clear.  For
example (and to note yet another worry regarding the anthropic
program), consider cosmologies with $Q\sim 0.01$ and $\Lambda\sim
10^{-114}$. The cosmology would (by TR's arguments) be dominated by
black holes, with (at best) extremely dense $10^{16}\msol$ galaxies
forming just after matter domination and just as $\Lambda-$domination
begins.  This cosmology would be rather inhospitable to life,
i.e. $\xi(Q\sim 0.01,\Lambda\sim 10^{-114}) \ll \xi(Q\sim
10^{-5},\Lambda\sim 10^{-124})$.  However, differences in $P$ might
compensate: consider a universe in which
$d^2P(\Lambda,Q^2)/d\Lambda\,dQ^2$ is flat (as per Weinberg's conjecture
applied to Linde's~\cite{lindebook} fiducial chaotic inflation model with
$V(\Lambda)=\lambda\phi^4/4$).  Then $P(Q\sim 0.01,\Lambda\sim
10^{-114}) \sim 10^{16} \xi(Q\sim 10^{-5},\Lambda\sim 10^{-124})$, and
such cosmologies must produce only one life-supporting star (perhaps
by some extremely baroque and unlikely process) per $10^{18}\msol$ of
baryons to compete with our cosmology.  It seems difficult to rule out
such a possibility.

\section{Summary and discussion}

	In \S~\ref{sec-acos} I denoted by
$\xi(\bar\alpha_1,..,\bar\alpha_N)$ the number of solar-mass,
metal-rich stars per baryon that would form over the lifetime of a
universe described by an FRW model specified by $N$ parameters $\alpha_i$
with values $\bar\alpha_i$.  If there is a probability
$P(\bar\alpha_1,..,\bar\alpha_N)$ that a given baryon finds itself in a
cosmology described by $\bar\alpha_i$, the probability ${\cal P}\equiv
P(\bar\alpha_1,..,\bar\alpha_N)\xi(\bar\alpha_1,..,\bar\alpha_N)$ should
describe the probability that a randomly chosen observer measures the
values $\bar\alpha_i$.  I then defined the `anthropic program' as the
computation of ${\cal P}$; if this probability distribution has a
single peak at a set $\bar\alpha_i^{\rm pk}$ {\em and} if these are
near the measured values $\bar\alpha_i^{\rm obs}$, then it could be
claimed that the anthropic program has `explained' the values
$\bar\alpha_i^{\rm obs}$ of the parameters of our cosmology by showing
that it is extremely likely for typical observers to measure values
very near $\bar\alpha_i^{\rm pk}\approx \bar\alpha_i^{\rm obs}$, so
that ({\em assuming} we are typical observers) it is not surprising that we
do.

	In \S~\ref{sec-cbb} I developed a class of cosmologies in
which many of the $\alpha_i$ take values quite different from those
deduced from observations.  Setting $\alpha_i =
\{\eta_\gamma,Q,\eta_{\rm dm},\eta_L,{\cal R},\Lambda\}$, the
observable universe can apparently be described by an FRW model with
$\bar\alpha_i^{\rm obs} =
\{\sim10^9,\sim10^{-5},\sim10,\ll10^9,\gg 10^{30},\lsim 10^{-56}\,{\rm
cm^{-2}}\}$. Observationally, we see that $\sim1\%$ of baryons form
stars, i.e.  $\xi(\bar\alpha_1^{\rm obs},..,\bar\alpha_N^{\rm
obs})\sim 0.01m_p/\msol$.  Let $\tilde\alpha_i \equiv
\bar\alpha_i/\bar\alpha_i^{\rm obs}$ (i.e. the the $\alpha_i$ in units
of their observed values).  Then in \S~\ref{sec-cbb} I argued that
similar values for $\xi$ can arise if
$\tilde\alpha_i\sim\{10^{-9},1,0,1,1,1\}$ if supermassive population
III objects form primordially and heat the cosmic medium up to
$\eta_\gamma \sim 10^9$.  Alternatively, solar-mass stars might form
primordially (with primordially formed metals), plausibly giving
similar values of $\xi$ for
$\tilde\alpha_i\sim\{10^{-9},10^{-4},0,1,1,1\}$ (if primordial clouds
fragment into $\msol$ stars) or
$\tilde\alpha_i\sim\{10^{-9},10^{-4},1,1,1,1\}$ (if not). Furthermore,
structure forms very early in such cosmologies, implying that $\xi$
can be high even in drastically different cosmologies such as (for
example) that described by
$\tilde\alpha_i\sim\{10^{-9},10^{-4},0,1,10^{5},10^{17}\}$.  Less
dramatically, but quite illustratively, the calculations of TR show
that one can construct a whole 4-dimensional region of parameter space
in which $\xi$ is within a few orders of magnitude, defined by
$\xi(y,x,1,1,w,z) \sim$(const) and parameterized by $x$, where
$10^{-3} \lsim x \lsim 100$, $0.1x^{7/8}\lsim y\lsim 10x^{3/4}$,
$w\gsim 0.1yx^{-1/2}$, and $z\lsim y^{-4}x^3\}$.

	At minimum, the existence of many independent maxima and
planes of degeneracy in $\xi$ -- widely separated in parameter space
-- should be discouraging for proponents of the anthropic program: it
implies that it is quite important to know the probabilities $P$,
which generally depend on poorly constrained models of the early
universe.  The hope that anthropic considerations would lead to only a
small allowed region of parameter space -- i.e. a small, sharp peak in
$\xi(\alpha)$ -- is not realized, and it seems that anthropic
arguments alone cannot simultaneously constrain multiple cosmological
parameters, even when many assumptions that are quite sympathetic to
the anthropic program are made.

	Drawing further conclusions from the arguments I have
presented requires assumptions about $P$.  Given strong {\em a priori}
confidence in a particular form of $P$, one could rule out the
anthropic program itself if that form of $P$ favors any of the
cosmologies I have developed here which are unlike our own; this might
imply, for example, that we simply happen to be `improbable' observers
and/or that there are {\em not} other universes in which the
parameters vary as assumed, or that it is misguided to consider as
possible only `observers' quite like ourselves. 
For those committed to the anthropic program,
the CBB cosmologies could be used constructively to constrain $P$: any
model in which $P$ is very strongly weighted toward one of the
`alternative' sets of $\bar\alpha_i$ is ruled out.

	This paper has largely addressed anthropic arguments
concerning cosmological parameters, but many of the issues raised
here clearly apply to anthropic arguments concerning more fundamental
constants.  The construction of specific counter-examples to these
arguments (i.e. cosmologies with very different fundamental constants
yet with observers), analogous to the CBB cosmologies developed here,
would seem to be an enormously more difficult technical task and would
require a much more careful assessment of what could reasonably
constitute an observer~\cite{obsarg}.  This difficulty would, however, be greatly
exceeded by the difficulty of rigorously arguing that {\em no} such
alternative cosmology exists.~\cite{bdm}

	 In conclusion, I have noted that one cannot simultaneously
anthropically explain the values of several parameters if the argument
for each parameter requires all others to be fixed.  It is possible,
then, to explain at most one parameter using such an argument.  To
explain more than one parameter, more than one parameter must be
varied among the ensemble members, and degeneracies can arise in the
probability distributions.  In the case that all six of the parameters
specifying the standard FRW cosmology are allowed to vary, I
find that it is possible to construct a cosmology in which all of the
parameters vary by (at least) several orders of magnitude from their
`observed' values, yet in which stars, planets, and intelligent life
can plausibly arise.  This greatly complicates, and reduces the
explanatory power of, anthropic arguments in cosmology.

\acknowledgements

I thank David Layzer, Eliot Quataert, Joop Schaye, Martin Rees, and Michael Dine for
helpful suggestions.  This work was supported by a grant from the
W.M. Keck foundation.


\begin{thebibliography}{99}
\bibitem{misceff} I have assumed that the number of independent
measurements in a sub-universe is proportional to the number of
suitable stars, neglected such effects as, for example,
high metallicity leading to more planets upon which multiple
civilizations might arise in a single solar system.  I have also
neglected factors like extinction-causing impacts, radioactivity-induced
plate tectonics, etc., which are possibly but not clearly necessary
for the evolution of observers.
\bibitem{obsarg} A proponent of the anthropic program could
self-consistently but unconvincingly argue that because we are human
(rather than some other sort of creature), we are justified in
assuming that human-like creatures dominate the set of observers
creatures making anthropic arguments.  How to define `human-like'
without making the argument circular is, however, unclear.
\end{thebibliography}
\end{document}